\documentclass[epj]{svjour}
\usepackage{psfig}
\begin{document}
\title{Analysis of the $\Lambda p$ Final State Interaction\\ 
in the 
Reaction $p+p\to K^+(\Lambda p)$.}
\author{F. Hinterberger\inst{1} \and A. Sibirtsev\inst{2,3}}   

\institute{Helmholtz-Institut f\"ur Strahlen- und Kernphysik, 
Universit\"at Bonn, Nu{\ss}allee 14-16, D-53115, Bonn, Germany
\and Institut f\"ur Kernphysik,   Forschungszentrum J\"ulich, 
D-52425 J\"ulich, Germany \and
Special Research Center for the Subatomic Structure of Matter (CSSM) 
and Department of Physics and Mathematical Physics, 
University of Adelaide, SA 5005, Australia  
}
\date{Received: date / Revised version: date}
%
\abstract{ 
The missing mass spectrum measured in high resolution studies of
the reaction $pp\to K^+X$ is analyzed with respect to the
strong final state interaction near the $\Lambda{p}$ production threshold. The
observed spectrum can be described by factorizing the reaction amplitude in
terms of a production amplitude and a final state scattering amplitude.
Parametrizing the $\Lambda{p}$ final state interaction in terms of
the inverse Jost function allows a direct extraction of the low-energy
phase-equivalent potential parameters. Constraints on the singlet and triplet
scattering lengths and effective ranges are deduced in a simultaneous fit of
the $\Lambda{p}$ invariant mass spectrum and the total
cross section data of the
free $\Lambda{p}$ scattering using the effective range approximation.
} 
\maketitle

\section{Introduction.}
Experimental information on the $\Lambda{p}$ interaction has been derived from
the analysis of hypernuclei, $\Lambda{p}$ scattering experiments and studies
of the $\Lambda{p}$ final state interaction (FSI) in strangeness transfer reactions.
The binding energy of light hypernuclei shows that the low-energy $\Lambda{p}$
interaction is attractive. In addition, the $\Lambda{p}$ interaction is
spin-dependent and the singlet interaction is stronger than the triplet
one~\cite{fhint:dal65,fhint:dal81,fhint:dov84}.  The free $\Lambda{p}$ scattering was studied in
bubble chamber measurements~\cite{fhint:ale68,fhint:sec68,fhint:ale69}.  In
the low momentum region the elastic cross sections were analyzed in terms of
the $S$-wave singlet and triplet scattering lengths and effective
ranges. However these determinations are characterized by large variances and
covariances since the data only support the determination of a spin-averaged
scattering length and effective range~\cite{fhint:ale69}.

The strong effect due to the $\Lambda{p}$ FSI was observed
in strangeness transfer reactions~\cite{fhint:tai69,fhint:pig77,fhint:pie88}
and and in associated strangeness production
reactions~\cite{fhint:mel65,fhint:ree68,fhint:hog68,fhint:sie94,fhint:bal98,fhint:bil98,fhint:mag01}. The
$\Lambda{p}$ production in the $K^-d{\to}\pi^-\Lambda{p}$,
$\pi^+d{\to}K^+\Lambda{p}$, $\gamma{d}{\to}K^0\Lambda{p}$ and
$pp{\to}K^+\Lambda{p}$ reactions can provide substantial improvement in an
evaluation of the low energy $\Lambda{p}$ scattering.

Intensive theoretical studies of the hyperon-nucleon interaction with the
Nijmegen~\cite{fhint:nag77,fhint:nag79,fhint:mae89,fhint:sto99} and
J\"ulich~\cite{fhint:hol89,fhint:reu94,Haidenbauer} potential models predict
the hyperon-nucleon potentials and phase shift parameters.  These models also
predict the sing\-let and triplet scattering length and effective range
parameters of the free $S$-wave $\Lambda{p}$ interaction.

The present paper refers to the associated strangeness reaction
$pp{\to}K^+(\Lambda{p})$ which is characterized by a strong FSI 
near the $\Lambda{p}$ production threshold.  In most
experiments the $\Lambda{p}$ system was measured inclusively.  
Exclusive measurements of the $pp{\to}K^+\Lambda{p}$ reaction were performed
at COSY~\cite{fhint:bal98,fhint:bil98} and
SATURNE~\cite{fhint:mag01}. Theoretical analyses of the reactions were
done~\cite{fhint:del89,fhint:lag91,Sibirtsev1,fhint:fae97,fhint:sib98,Sibirtsev2,Tsushima1,Sibirtsev3,Shyam1,Gasparian1,fhint:kel00,Shyam2,Gasparian2,Gasparian3}
by applying the meson exchange model and including the $\Lambda{p}$
FSI generally modelled by Nijmegen or J\"ulich potentials.

The aim of the present paper is to perform an analysis of the $\Lambda{p}$ FSI
in the reaction $pp{\to}K^+(\Lambda{p})$ and an evaluation of the $\Lambda{p}$
low energy interaction parameters. Using the Watson-Migdal
approximation~\cite{wat52,mig55,gol64,Gillespie} the reaction amplitude is factorized in
terms of a production matrix element and a FSI enhancement factor, which can be
represented by the inverse Jost function~\cite{jos47,jos52,bar49a,bar49b}. We
analyze experimental results from SATURNE collected by Siebert et
al.~\cite{fhint:sie94}, which are characterized by a high statistical accuracy
and a high invariant mass resolution. Furthermore, in the fitting procedure
the missing mass resolution is taken into account by folding the theoretical
expressions with the experimental resolution function.

The analysis shows that the shape of the sharply rising invariant mass
spectrum depends strongly on the singlet and triplet scattering length and
effective range parameters. But only two parameters, the spin-averaged
scattering length and effective range parameters, can be deduced within an
acceptable confidence level by fitting the $\Lambda{p}$ missing mass
spectrum.  Additional information can be obtained by taking the total cross
section data for the free $\Lambda{p}$ scattering into account and including
these data in an overall fit.  At low energies the total cross section can be
described in a model-independent way using the effective range
approximation~\cite{fhint:sch47,fhint:bet49,fhint:bla54}.  Thus, by fitting
simultaneously the $\Lambda{p}$ invariant mass spectrum and the available total cross
section data of the free $\Lambda{p}$ scattering severe constraints
on the singlet and triplet scattering length and effective range parameters
can be deduced.  This method allows also to test theoretical model
predictions.

\section{The formalism.}
\subsection{Phase space distribution.}
The $pp{\to}K^+\Lambda{p}$ double differential cross section is given as
\begin{equation}
\frac{d^2\sigma}{d\Omega_{K}dM_{\Lambda{p}}}= |{\tilde{\cal M}|^2} \, \Phi_3 ,
\end{equation}
where ${\tilde {\cal M}}$ is the Lorentz-invariant reaction amplitude and the 
three-body phase space distribution function is 
\begin{equation}
\Phi_3{=} \frac{\pi}{16(2\pi)^5} \frac{p_{K}^2 q}{p_p m_p
[(E_{p}+m_p)p_{K}-E_{K}p_{p}\cos\theta_{K}]},
\label{phase}
\end{equation}
where $q$ is the momentum of $\Lambda$ in the Gottfried-Jackson rest-frame of the
produced two-particle subsystem $X{=}\Lambda{+}p$,  $M_{\Lambda{p}}$ is the
corresponding invariant mass and $p_p$, $E_p$, $p_K$, $E_K$, $\theta_K$,
$\Omega_K$ are defined in the laboratory system. Obviously, in inclusive
measurements the invariant mass $M_{\Lambda{p}}$ is equal to the missing mass
$M_X$ below the $\Sigma$-hyperon production threshold.  Eq.(\ref{phase}) is
consistent with the kinematical definitions of
Refs.~\cite{fhint:byc73,fhint:pdg98}.

\subsection{Final state interaction.}
In the Watson-Migdal approximation~\cite{wat52,mig55,gol64} the FSI is taken
into account by introducing a FSI enhancement factor $|C_{FSI}|^2$,
\begin{equation}
\frac{d^2\sigma}{d\Omega_{K} dM_{\Lambda{p}}}= |{\cal M}|^2 \,
|C_{FSI}|^2 \, \Phi_3,
\label{watson}
\end{equation}
where now ${\cal M}$ is pure production matrix element and 
the FSI amplitude $C_{FSI}$ depends on the internal momentum $q$ of the
$\Lambda{p}$ subsystem. It converges to 1 for $q{\to}\infty$ where the
$S$-wave FSI enhancement vanishes.

Applying the factorization we assume that the production
operator ${\cal M}$ is constant, i.e. does not depend 
on the internal kinetic energy of the $\Lambda{p}$ subsystem.
In case of the $pp{\to}K^+\Lambda{p}$
reaction this assumption is supported by the kinematics which provides
a focus onto the $\Lambda{p}$ FSI. The internal kinetic energy
of the $\Lambda{p}$ subsystem is almost zero 
near the $\Lambda{p}$ threshold whereas
the $K^+ \Lambda$- and $K^+{p}$-subsystems have
large internal kinetic energies.
Even if the $pp{\to}K^+\Lambda{p}$ reaction is dominated
\cite{Sibirtsev2,Tsushima1,Sibirtsev3,Shyam1} by 
intermediate baryonic resonances coupled to
the $K^+\Lambda$ system a small variation of
the invariant $\Lambda{p}$ mass does practically not affect
the production amplitude.


The methods for studying the FSI between the particles have
been developed in different areas of physics, ranging
from atomic physics to high energy particle physics \cite{Gillespie}. 
Taking the inverse Jost function~\cite{jos47,jos52} the correction due to the
FSI is given as 
\begin{equation}
C_{FSI}=\frac{q-i\beta}{q+i\alpha},\; \; \;
|C_{FSI}|^2=\frac{q^2+\beta^2}{q^2+\alpha^2}.
\end{equation} 
The potential parameters $\alpha$ and $\beta$ can be used to establish
phase-equivalent Bargmann potentials~\cite{bar49a,bar49b}.  They are related
to the scattering lengths $a$, and effective ranges $r$ of the low-energy
$S$-wave scattering
\begin{equation}
\alpha=\frac{1}{r}\left(1-\sqrt{1-2\frac{r}{a}}\right),\; \; \; \beta
=\frac{1}{r}\left(1+\sqrt{1-2\frac{r}{a}}\right).
\end{equation}

The $\Lambda{p}$ system can couple to singlet $^1S_0$ and triplet $^3S_1$
states. Near production threshold the singlet-triplet transitions due to the
final state interaction cannot occur. Therefore, the contributions of the
spin-singlet and spin-triplet final states can be added incoherently.  Taking
the spin-statistical weights into account the unpolarized double differential
cross section may be written as
\begin{eqnarray}
\frac{d^2\sigma}{d\Omega_{K}dM_{\Lambda{p}}}= \Phi_3 \left[\, 0.25 \, |{\cal
M}_s|^2 \, \frac{q^2+\beta_s^2}{q^2+\alpha_s^2} \right. \nonumber \\
{+}\left.\, 0.75 \, |{\cal M}_t|^2 \,
\frac{q^2+\beta_t^2}{q^2+\alpha_t^2}\right].
\label{miss}
\end{eqnarray}
This equation leaves six free parameters, the singlet and triplet potential
parameters $\alpha_s$, $\beta_s$, $\alpha_t$, $\beta_t$ and the production
matrix elements $|{\cal M}_s|$ and $|{\cal M}_t|$.  Instead of the parameters
$\alpha_s$, $\beta_s$, $\alpha_t$ and $\beta_t$ one can equally well use the
singlet and triplet scattering length and effective range parameters $a_s$,
$r_s$, $a_t$ and $r_t$.  The functional dependence on the invariant mass
$M_{\Lambda{p}}$ can be evaluated by inserting the corresponding expression for the
internal momentum $q$ of the $\Lambda{p}$ system,
\begin{equation}
q=\frac{\sqrt{M_{\Lambda{p}}^2-(m_\Lambda+m_p)^2}
\sqrt{M_{\Lambda{p}}^2-(m_\Lambda-m_p)^2}} {2M_{\Lambda{p}}}.
\end{equation}

\subsection{Missing mass resolution.}
The theoretical missing mass spectrum of Eq.(\ref{miss}) has to be folded with
the missing mass resolution function before comparing with the
data. Introducing the missing mass resolution function $g({\tilde
M}_{\Lambda{p}}{-}M_{\Lambda{p}})$ and denoting the r.h.s. of Eq.(\ref{miss})
by $f_0(M_{\Lambda{p}})$ yields the distribution function $f(M_{\Lambda{p}})$
which has to be compared with the data,
\begin{equation}
f(M_{\Lambda{p}})=\int \!\!f_0({\tilde M}_{\Lambda{p}}) \, g({\tilde
M}_{\Lambda{p}}{-}M_{\Lambda{p}}) \, d{\tilde M}_{\Lambda{p}}
\label{res}
\end{equation}
where the resolution function is given as
\begin{equation}
g({\tilde M}_{\Lambda{p}}{-}M_{\Lambda{p}}){=} \frac{1}{\sqrt
{2\pi}}\frac{1}{\sigma_M} \exp-\frac{({\tilde
M}_{\Lambda{p}}{-}M_{\Lambda{p}}{+} \Delta M_{\Lambda{p}})^2}{2\sigma_{M}^2}.
\end{equation}
Here, $\sigma_{M}$ denotes the one standard deviation width and the shift
$\Delta M_{\Lambda{p}}$ takes a systematic calibration error of the invariant
mass scale into account.  For the comparison with the SATURNE data the
values $\Delta M_{\Lambda{p}}{=}1.7$~MeV and 
$\sigma_M{=}2.0$~MeV \cite{fhint:sie94} are used.

\subsection{The  $\Lambda{p}$ cross section.} 
In the effective range
approximation~\cite{fhint:sch47,fhint:bet49,fhint:bla54} 
the total
cross section of the free $\Lambda{p}$ elastic scattering can be expressed in
terms of the singlet and triplet scattering length and effective range
parameters $a_s$, $a_t$, $r_s$ and $r_t$,
\begin{equation}
\sigma_{\Lambda{p}}{=} \frac{\pi}{q^2{+}\left(-\frac{1}{a_s}{+}\frac{r_s
q^2}{2}\right)^2} {+} \frac{3\pi}{q^2{+}\left(-\frac{1}{a_t}{+}\frac{r_t
q^2}{2}\right)^2}.
\label{cross}
\end{equation}
Here, $q$ is the cm-momentum of the $\Lambda{p}$ scattering.  This equation
can be applied at low energies where $S$-wave contributions dominate.

\section{The results of the fits.}
The program Minuit of the CERN program library~\cite{fhint:cer94} was used in
order to perform nonlinear least-square fits.  
We fit the experimental results ~\cite{fhint:sie94} for missing mass spectrum from
$pp{\to}K^+X$ reaction measured at proton beam energy $T_p{=}2.3$~GeV and kaon
emission angle of $\theta_K{=}10.3^\circ$ using
Eqs.(\ref{miss}) and (\ref{res}), respectively. 
We fit the strong FSI enhancement near the $\Lambda{p}$
threshold assuming that the $S$-wave FSI is dominant.
Therefore, we take only data corresponding
to very low c.m. momenta  $q$ in the $\Lambda{p}$ system.
The fits include 29 experimental points of the missing
mass spectrum from the $\Lambda{p}$ threshold at 2054~MeV up to the invariant
mass of the $\Lambda{p}$ system of 2095~MeV.  This $M_{\Lambda{p}}$
range corresponds to c.m.  momenta $q$ in the $\Lambda{p}$ system between 0
and 200 MeV/c.

Low energy total $\Lambda{p}$
cross section data from bubble chamber
measurements~\cite{fhint:ale68,fhint:sec68,fhint:ale69} are fitted  
simultaneously using
Eq.(\ref{cross}). The fits include 12 total
$\Lambda{p}$ cross section data points covering c.m. momentum range
60${\le}q{\le}$135~MeV/c. 


It is worthwhile to mention that the  $pp{\to}K^+X$ reaction
provides high quality data even at very small $q$ near the
$\Lambda p$ threshold whereas $\Lambda{p}{\to}\Lambda{p}$ data
are available only for $q{\geq}60$~MeV/c. 
Also the number of data and the accuracy of the $pp{\to}K^+X$
measurement is substantially higher.
Therefore it is very attractive to include the $pp{\to}K^+X$ data 
in the evaluation of the  low-energy phase-equivalent 
potential parameters as far as the evaluation  remains model
independent. 

The quality of the least-square fits is given by total $\chi^2$ and reduced
$\chi^2/ndf$ where $ndf$ is the number of degrees of freedom given by the
number of data minus the number of the fit parameter. In calculating the
parameter errors nonlinearities and parameter correlations were taken into
account. The error for a given parameter is defined as the change of that
parameter which causes $\chi^2$ to increase by one while re-fitting all other
free parameters.  Due to the nonlinearities the resulting error intervals are
in general asymmetric.

\subsection{Three-parameter fit.}

In a three-parameter fit we determine  spin-averaged parameters 
by applying the 
constraints
\[
|{\cal M}_s|^2{=}|{\cal M}_t|^2{=}|\bar{{\cal M}}|^2,\; \; 
a_s{=}a_t{=}\bar{a},\; \; r_s{=}r_t{=}\bar{r}.
\]
The three-parameter fit yields the production matrix element
squared $|\bar{\cal M}|^2$ and the spin-averaged scattering length $\bar{a}$
and effective range $\bar{r}$.

\begin{figure}[h!]\vspace*{-3mm}
\begin{center}
\psfig{file=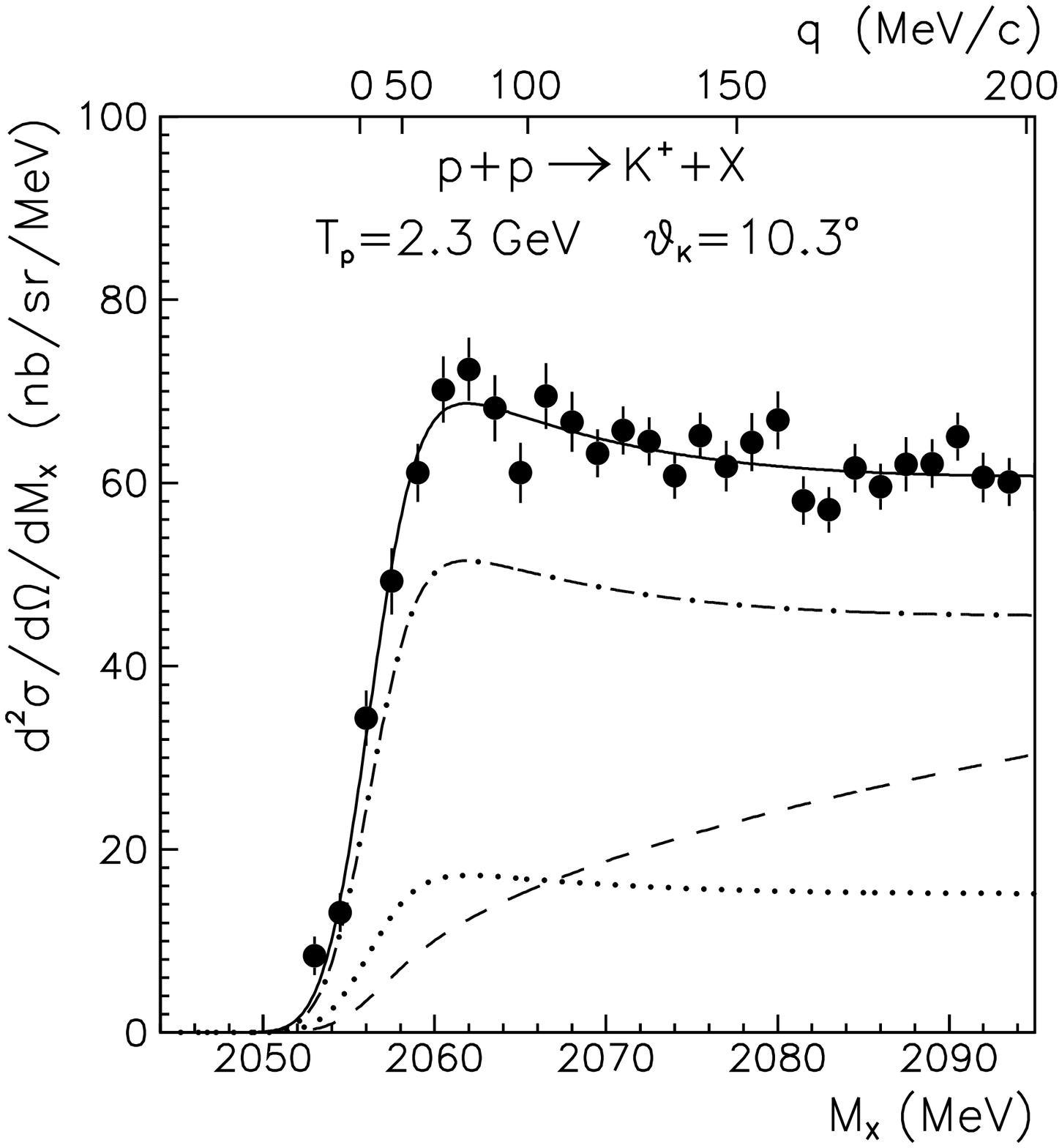,width=9cm,height=8cm}\vspace*{-10mm}
\psfig{file=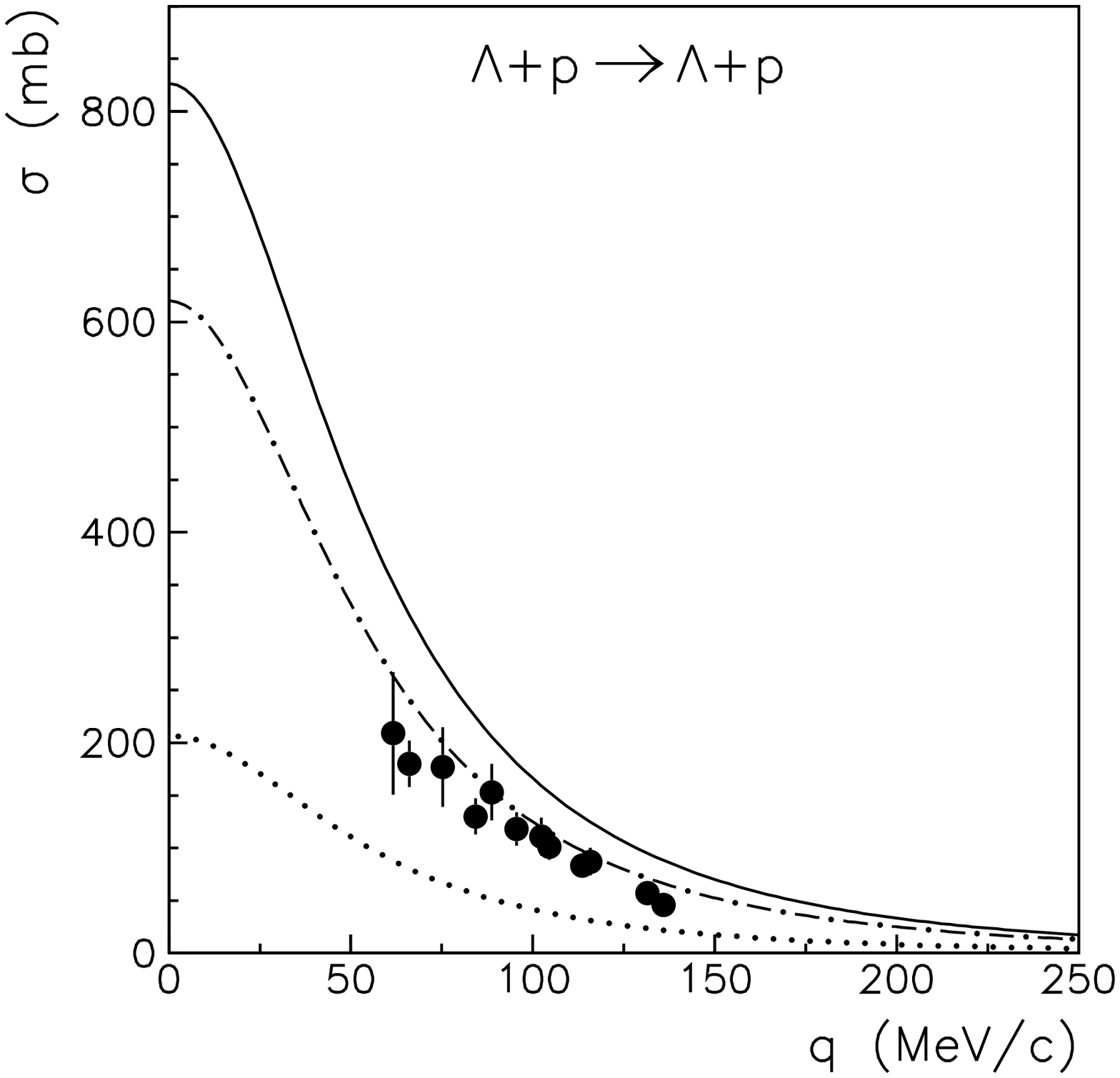,width=9cm,height=8cm}\vspace*{-3mm}
\end{center}
\caption{Top: Missing mass spectrum  of the reaction $pp{\to}K^+X$
 measured at $T_p{=}2.3~GeV$
and  $\theta_K{=}10.3^\circ$ \cite{fhint:sie94}. 
The upper axis indicates
the c.m.  momentum $q$ of the $\Lambda{p}$ system. 
Bottom: Total $\Lambda{p}$ cross 
section\cite{fhint:ale68,fhint:sec68,fhint:ale69} as a function of the
c.m. momentum $q$. 
Solid lines: Fit curves with parameters given by
Eq.(\ref{fitmm3cross0}) from a 
three-parameter fit of the missing mass spectrum alone, 
dashed line: phase space distribution,
dotted lines: singlet contributions,
dashed-dotted lines: triplet contributions.}
\label{mm3cross0}
\end{figure}

In a first step we fit only the missing mass spectrum
without taking the $\Lambda p$  total cross section data into account.
The three-parameter fit yields an excellent description
of the missing mass spectrum 
but fails completely to reproduce the total
cross section data ($\chi^2/ndf{=}18.0$, see Fig.~\ref{mm3cross0}).
The fit parameters of the missing mass spectrum are
\begin{eqnarray}
|\bar{\cal M}|^2&=&15.4_{-1.6}^{+1.5}\;{\rm b/sr},\; \chi^2/ndf=0.98 \nonumber \\
\bar{a}&=&-2.57_{-0.23}^{+0.20}\; {\rm fm},\; 
\bar{r}=2.47_{-0.24}^{+0.23}\; {\rm fm}.  
\label{fitmm3cross0}
\end{eqnarray}
The dotted and dashed-dotted lines in Fig. \ref{mm3cross0}
and in the following Figs. \ref{mm1cross2} - \ref{mm5cross2}
show the corresponding singlet and triplet contributions.
The dashed lines show the phase space distributions
without the FSI enhancement factor, i.e.
$(0.25|{\cal M}_s|^2 + 0.75|{\cal M}_t|^2)\Phi_3$.

Vice versa, we take only
the total cross section data into account and
determine $\bar{a}$ and $\bar{r}$ in a two-parameter fit.
The resulting fit parameters are
\begin{equation}
\bar{a}=-1.81_{-0.21}^{+0.18}\; {\rm fm},\; 
\bar{r}=3.24_{-0.48}^{+0.48}\; {\rm fm},\; 
\chi^2/ndf=0.39.
\label{cross2}
\end{equation}
Taking those parameters fixed and fitting only $|\bar{\cal M}|^2$
yields for the missing mass spectrum
\begin{equation}
|\bar{\cal M}|^2=19.5_{-0.2}^{+0.2}\;{\rm b/sr},\; 
\chi^2/ndf=3.7.
\label{fitmm1cross2}
\end{equation}
This procedure yields an excellent fit
of the total cross section data but fails completely to describe the
missing mass spectrum (see Fig.~\ref{mm1cross2}).
\begin{figure}[h!]\vspace*{-3mm}
\begin{center}
\psfig{file=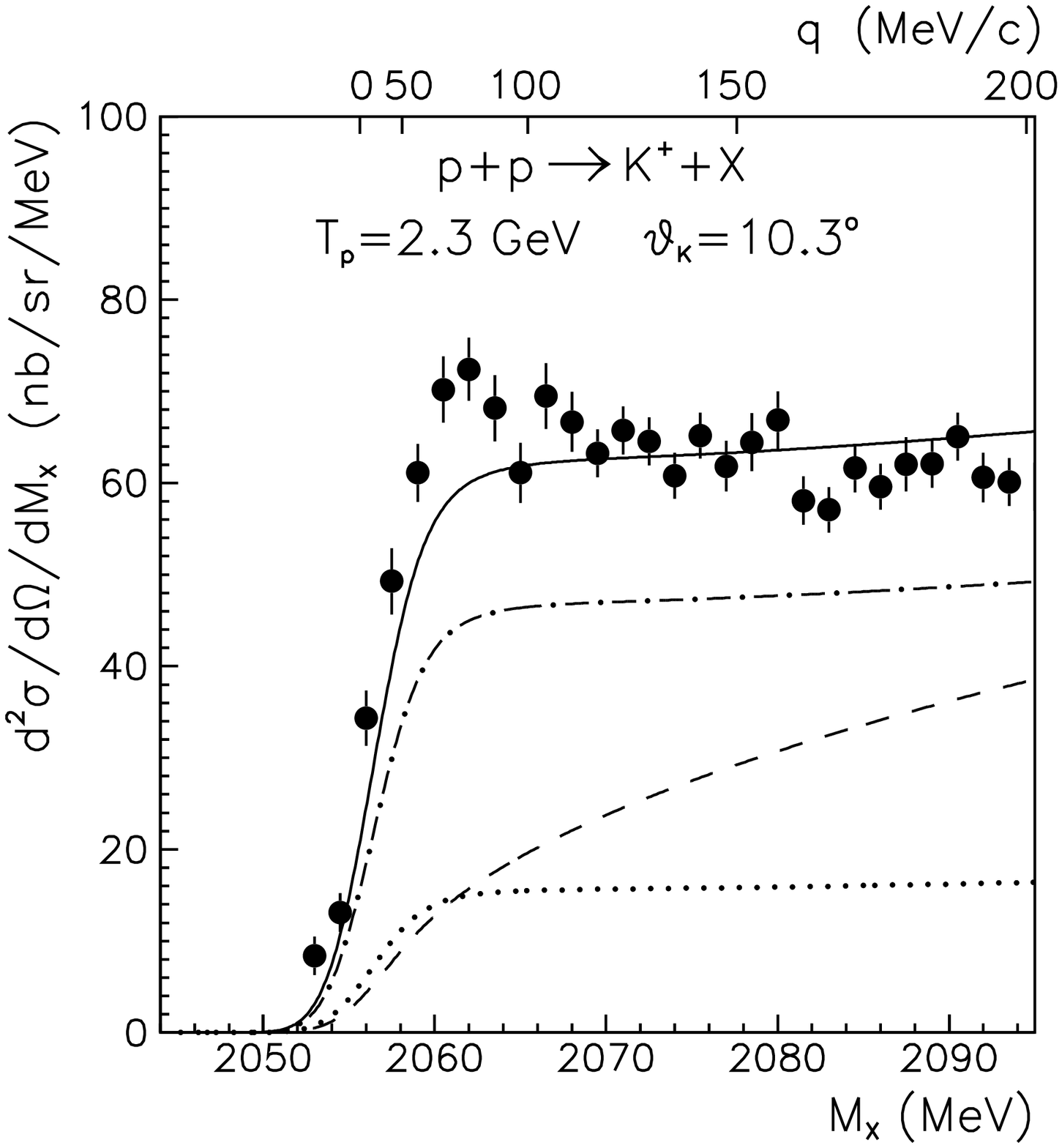,width=9cm,height=8cm}\vspace*{-10mm}
\psfig{file=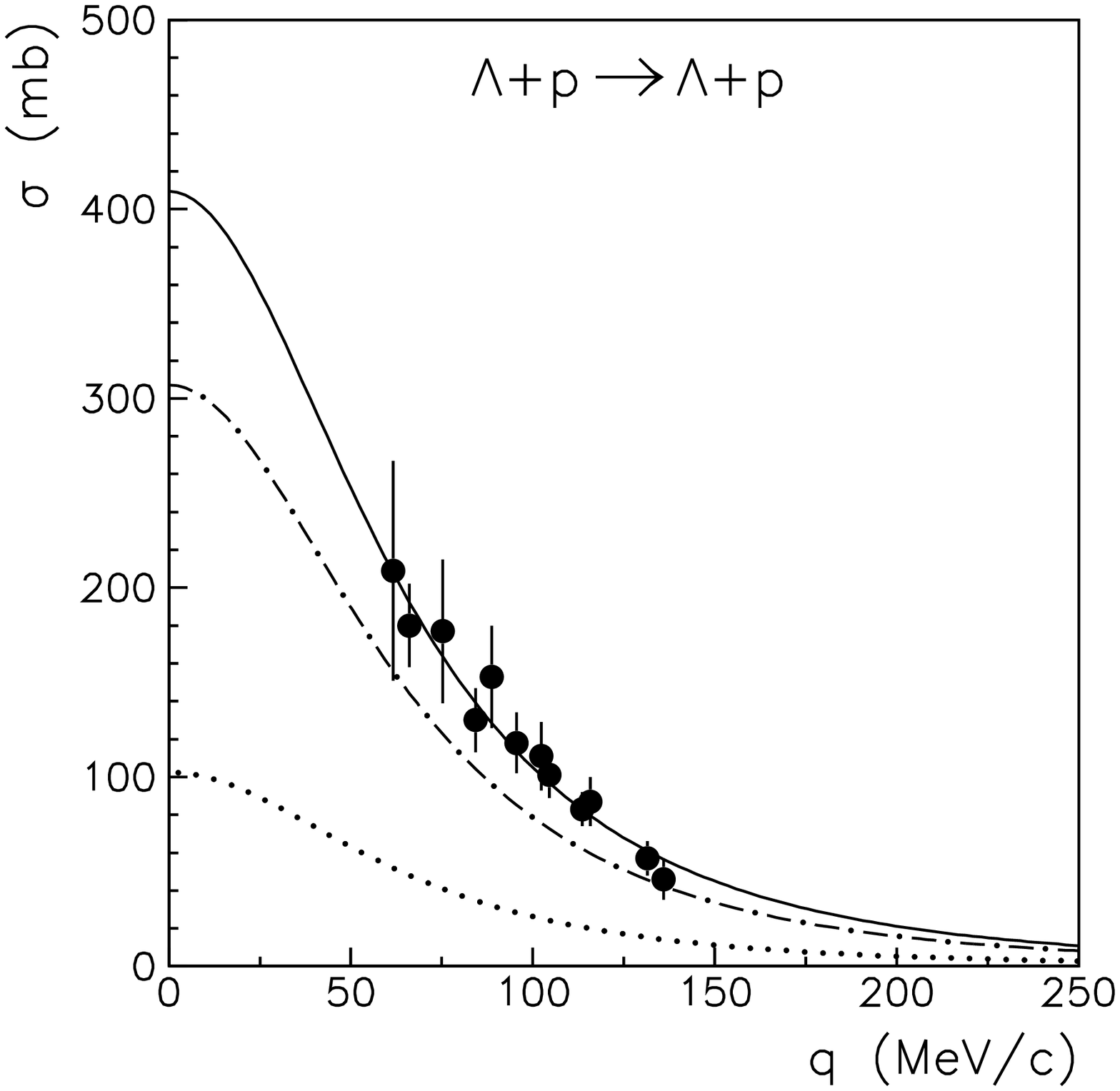,width=9cm,height=8cm}\vspace*{-3mm}
\end{center}
\caption{
Same as Fig.~\ref{mm3cross0}.
Solid lines: Fit curves with $\bar{a}$ and $\bar{r}$ given by
Eq.~(\ref{cross2}) from a 
two-parameter fit of the total cross section data alone
and  $|\bar{\cal M}|^2$ given by Eq.~(\ref{fitmm1cross2}), 
dashed line: phase space distribution,
dotted lines: singlet contributions,
dashed-dotted lines: triplet contributions.}
\label{mm1cross2}
\end{figure}

In a next step, we determine spin-averaged parameters
in a combined fit, i.e. by fitting simultaneously
the missing mass spectrum and the total cross section data.
The resulting parameters are
\begin{eqnarray}
|\bar{\cal M}|^2&=&16.9_{-1.2}^{+1.2}\;{\rm b/sr},\; 
\chi^2/ndf=2.2, \nonumber \\
\bar{a}&=&-1.91_{-0.11}^{+0.10}\; {\rm fm},\; 
\bar{r}=2.74_{-0.20}^{+0.20}\; {\rm fm}.
\label{fitmm3cross2}
\end{eqnarray}
This procedure fails to  describe both,
the missing mass spectrum and 
the total cross section data (see Fig.~\ref{mm3cross2}).
This failure is a direct indication that the spin dependence
of the $\Lambda{p}$ interaction must be taken into account.

\begin{figure}[h!]\vspace*{-3mm}
\begin{center}
\psfig{file=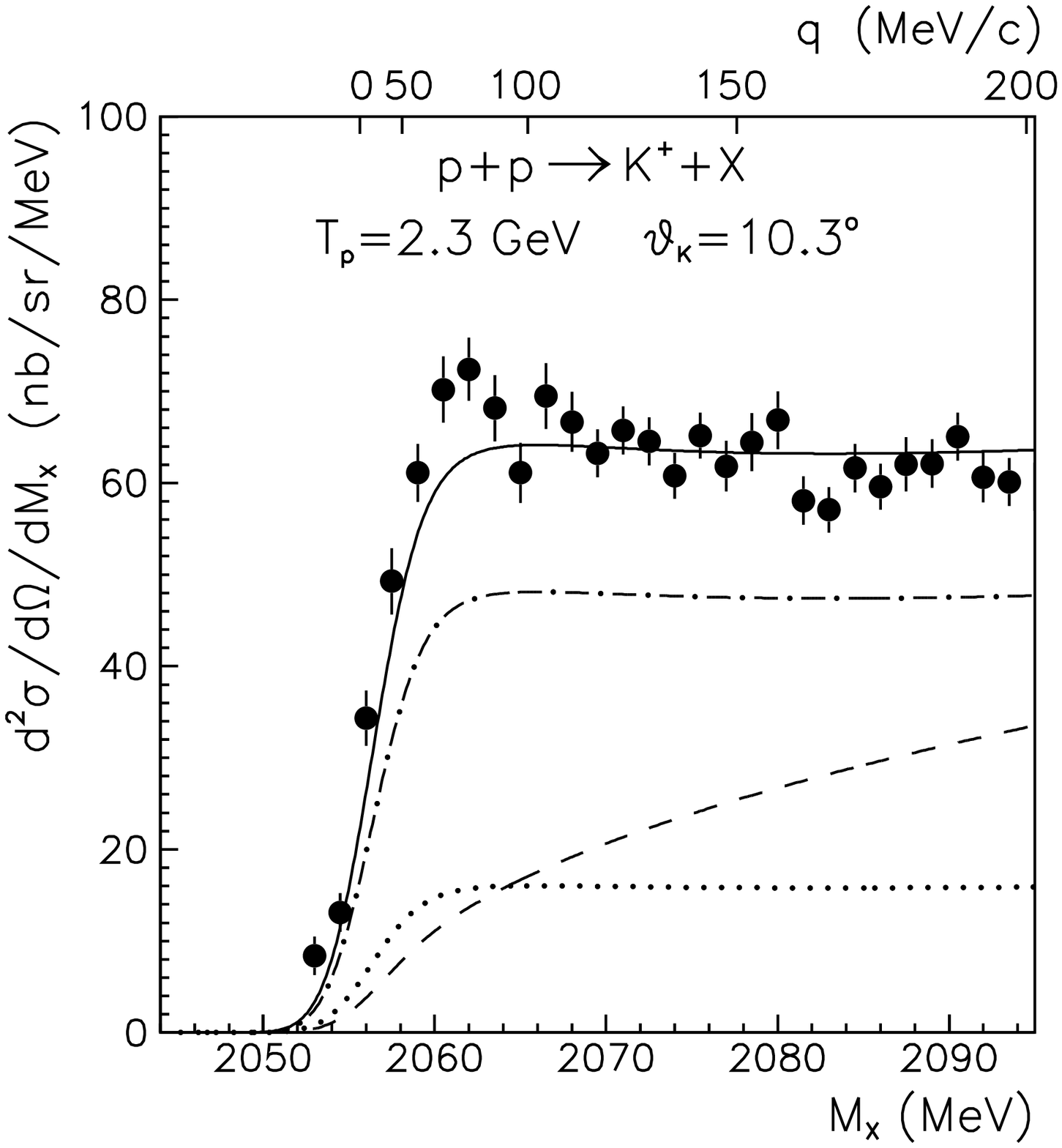,width=9cm,height=8cm}\vspace*{-10mm}
\psfig{file=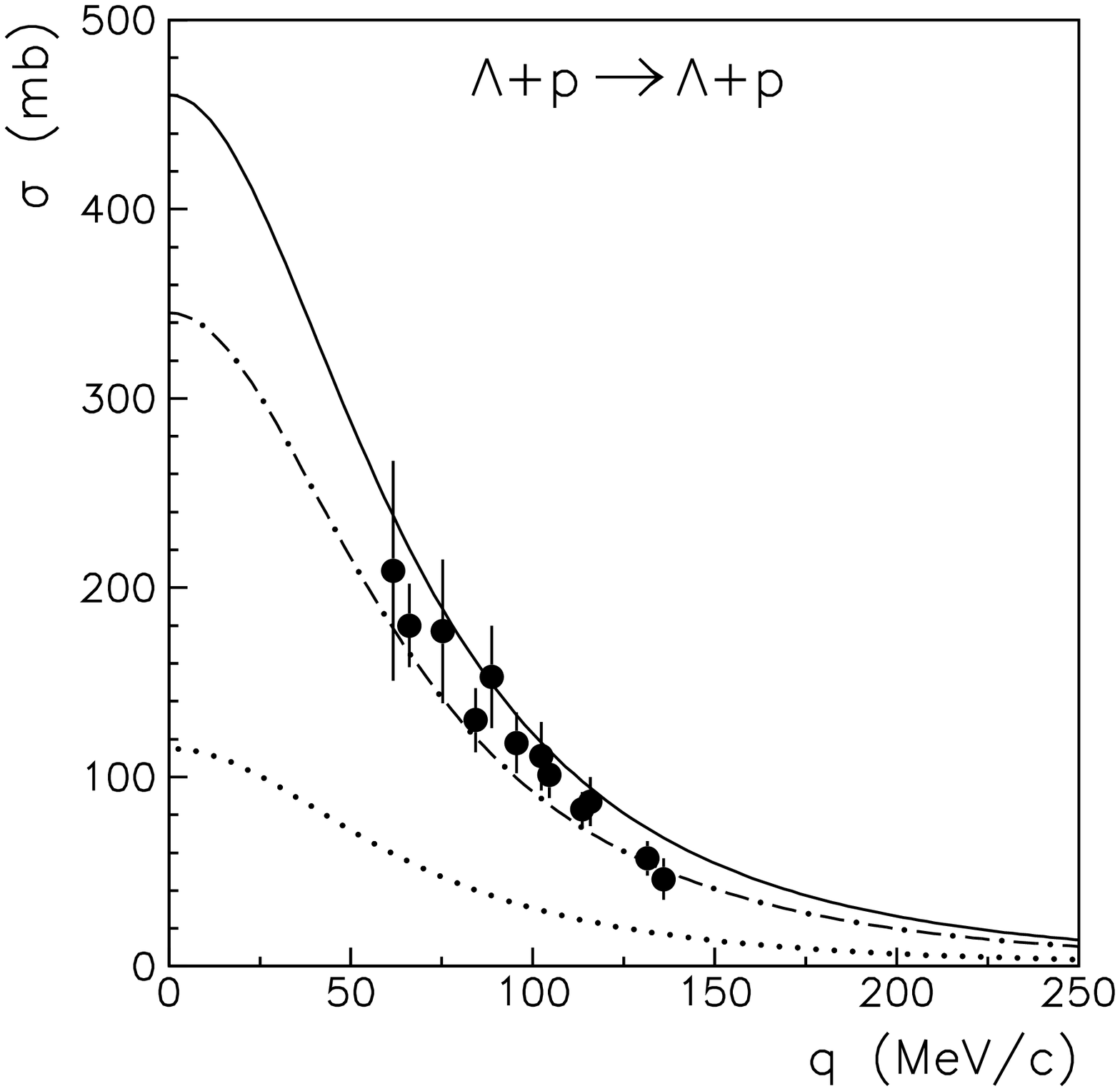,width=9cm,height=8cm}\vspace*{-3mm}
\end{center}
\vspace*{-5mm}
\caption{
Same as Fig.~\ref{mm3cross0}.
Solid lines: Fit curves with parameters given by
Eq.(\ref{fitmm3cross2}) from a combined  
three-parameter fit of the missing mass spectrum and the 
total cross section data, 
dashed line: phase space distribution,
dotted lines: singlet contributions,
dashed-dotted lines: triplet contributions.}
\label{mm3cross2}
\end{figure}

\begin{figure}[h!]
\vspace*{-3mm}
\begin{center}
\psfig{file=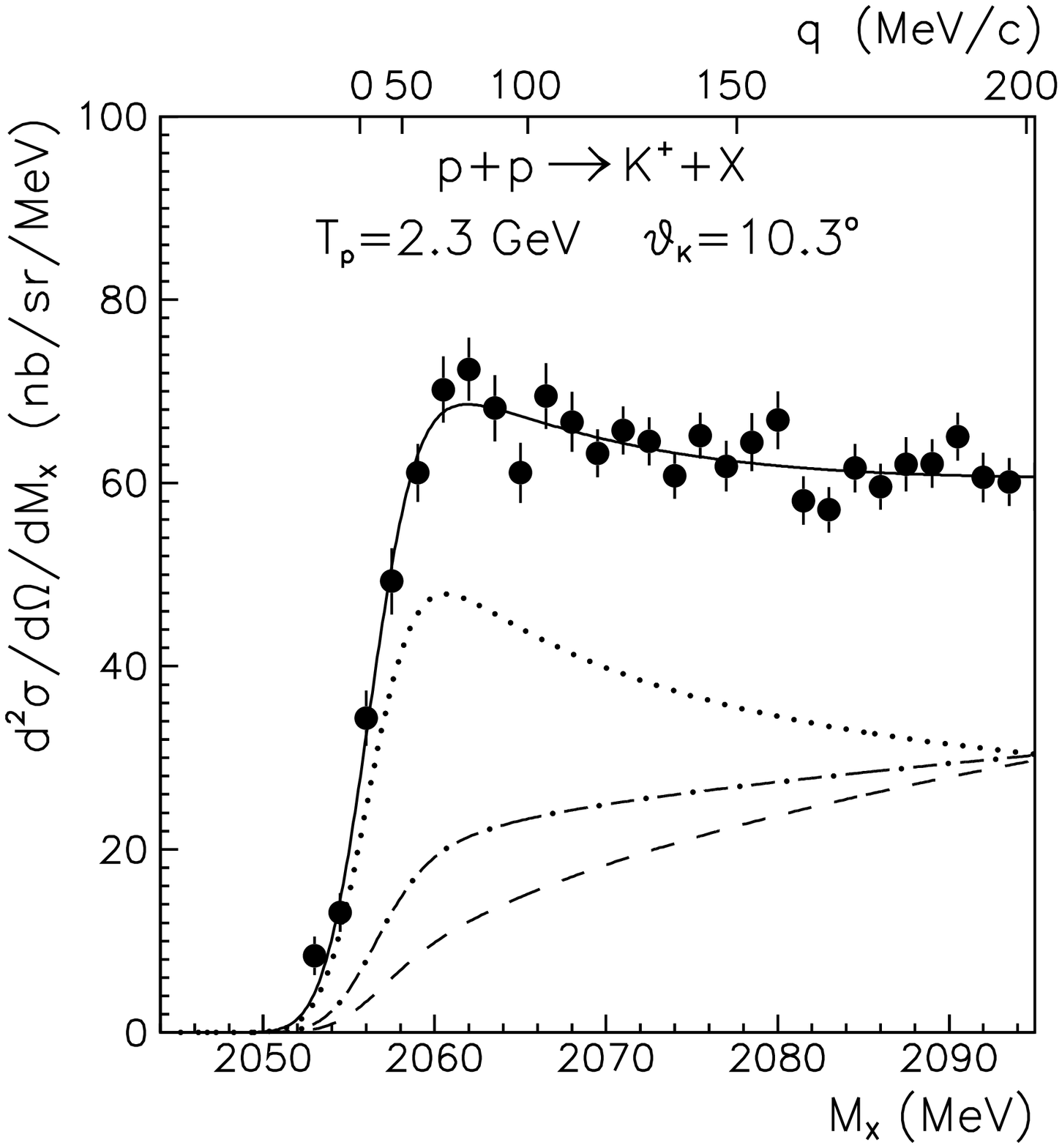,width=9cm,height=8cm}\vspace*{-10mm}
\psfig{file=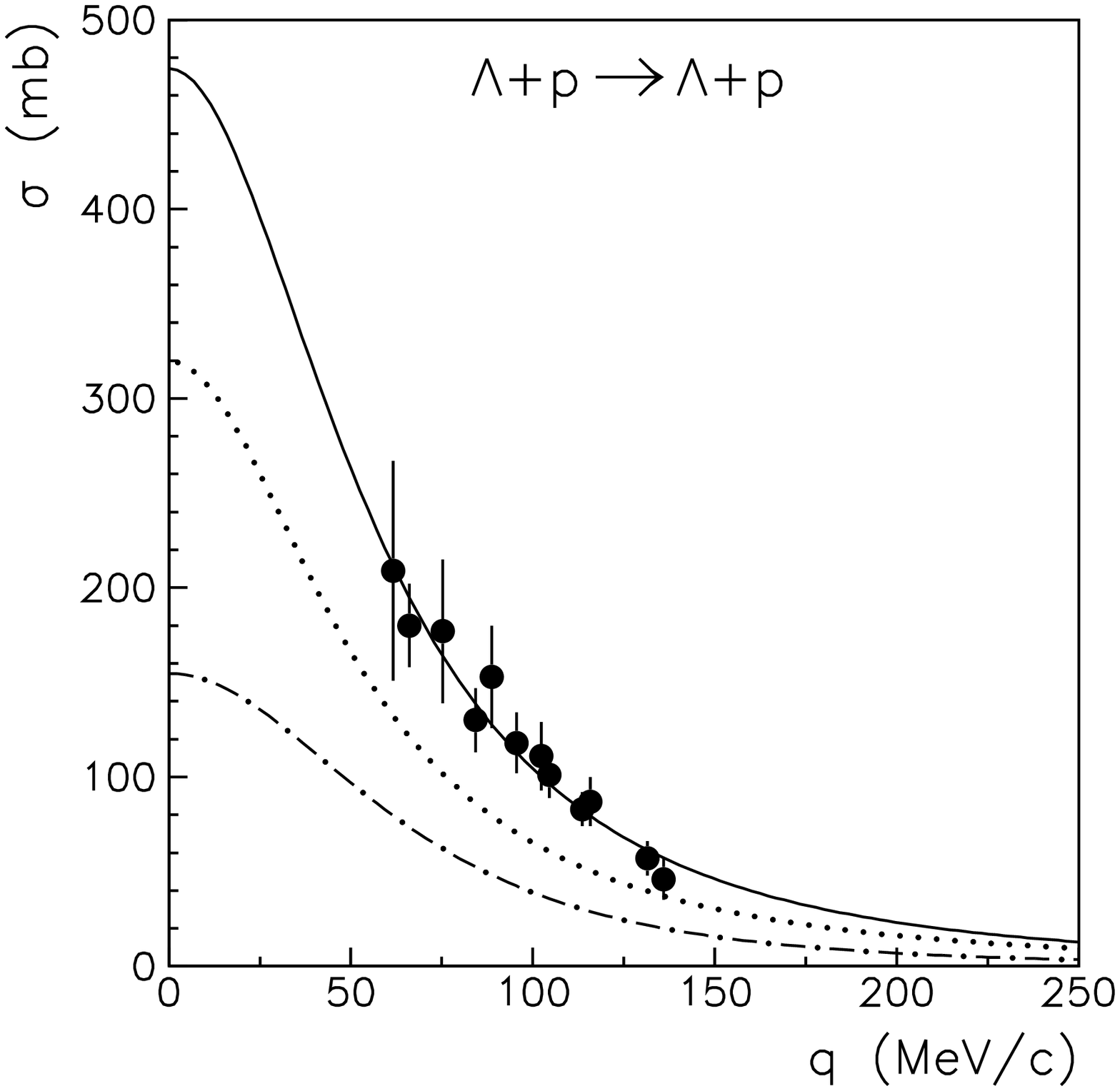,width=9cm,height=8cm}\vspace*{-3mm}
\end{center}
\vspace*{-5mm}
\caption{
Same as Fig.~\ref{mm3cross0}.
Solid lines: Fit curves with parameters given by
Eq.(\ref{fitmm5cross2}) from a combined  
five-parameter fit of the missing mass spectrum and the 
total cross section data, 
dashed line: phase space distribution,
dotted lines: singlet contributions,
dashed-dotted lines: triplet contributions.}
\label{mm5cross2}
\end{figure}

\subsection{Five-parameter fit.}

Now the data on total  $\Lambda{p}$ cross section and $pp{\to}K^+X$
missing mass spectrum are fitted in a combined fit
with the singlet and triplet scattering lengths and effective ran\-ges
$a_s$, $r_s$, $a_t$, $r_t$ 
as separate free parameters.
Taking the unknown quantities  
$|{\cal M}_s|^2$ and $|{\cal M}_t|^2$
into account
a six-parameter fit should be performed.
However, it turned out that the $\chi^2$-criterion cannot be
used to determine simultaneously $|{\cal M}_s|^2$ and $|{\cal M}_t|^2$.
This is due to the fact that the resulting $\chi^2$ 
depends only weakly on the ratio
$|{\cal M}_t|^2{/}|{\cal M}_s|^2$ as is indicated in
Table~\ref{tab1}. 
Therefore, five-parameter fits were performed
taking  $|{\cal M}_s|^2$ as  free parameter and 
the ratio $|{\cal M}_t|^2/|{\cal M}_s|^2$
as  fixed parameter.  
By this method valuable
constraints on the
singlet and triplet scattering lengths
and effective ranges can be deduced from the data.

It should be mentioned that the spin-statistical weights 0.25 and 0.75 of the
singlet and triplet contributions have already been taken
into account in the theoretical ansatz (\ref{miss}).
Therefore the five-parameter search was started with the constraint
$|{\cal M}_t|^2/|{\cal M}_s|^2{=}1$ yielding
a solution with $\chi^2/ndf{=}0.82$.
The resulting parameters are
\begin{eqnarray}
|{\cal M}_s|^2&=&15.0_{-1.6}^{+1.4}\;{\rm b/sr},\; 
\chi^2/ndf=0.82, \nonumber \\
a_s&=&-3.2_{-0.6}^{+0.4}\; {\rm fm},\; 
r_s=1.25_{-0.15}^{+0.13}\; {\rm fm}, \nonumber \\
a_t&=&-1.3_{-0.5}^{+0.4}\; {\rm fm},\; 
r_t=5.4_{-1.6}^{+1.6}\; {\rm fm}.
\label{fitmm5cross2}
\end{eqnarray}
Both,  the missing mass spectrum and
the total cross section data (see Fig.\ref{mm5cross2})
are perfectly reproduced.
Then, the ratio $|{\cal M}_t|^2/|{\cal M}_s|^2$
was varied over a wide range  of values
and best fit solutions of similar quality with 
$\chi^2/ndf$ varying between 0.80 and 0.83 were found.
The resulting  fit parameters are listed in 
Table~\ref{tab1}
for $0{\leq} |{\cal M}_t|^2/|{\cal M}_s|^2{\leq} 8$.

A characteristic feature of the best fit solutions
is the fact that the singlet FSI enhancement factor at $q{=}0$
is much larger than the triplet one,
e.g. $\beta_s^2/\alpha_s^2{=}49.1$ and  $\beta_t^2/\alpha_t^2{=}3.9$
for $|{\cal M}_t|^2/|{\cal M}_s|^2{=}1$.
In spite of the statistical weight 0.25 the 
singlet contribution dominates the
${\rm p}{\rm p}{\to}{\rm K}^+ \Lambda {\rm p}$ cross section 
at $q{=}0$,
e.g. $0.25|{\cal M}_s|^2\beta_s^2/\alpha_s^2{=}183$ b/sr
and $0.75|{\cal M}_t|^2\beta_t^2/\alpha_t^2{=}43.5$~b/sr
for $|{\cal M}_t|^2/|{\cal M}_s|^2{=}1$.
Solutions where  the triplet contribution is
larger than the singlet contribution  
do not fit the FSI enhancement of the
missing mass spectrum near $q=0$
and can be excluded on the
basis of the $\chi^2$ criterion. 
This holds true even if one varies the ratio
$|{\cal M}_t|^2/|{\cal M}_s|^2$ in a wide range.

\begin{figure}[b]\vspace*{-7mm}
\hspace*{-1mm}\psfig{file=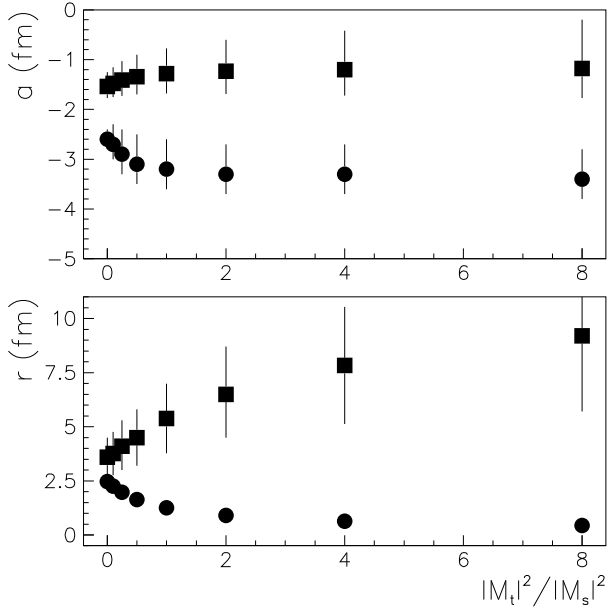,width=9.2cm,height=9.2cm}\vspace*{-3mm}
\caption{The singlet (circles) and triplet (squares) scattering lengths and
effective ranges as a function of the ratio $|{\cal M}_t|^2/|{\cal M}_s|^2$
resulting from the overall five-parameter fit to the total $\Lambda{p}$ cross
section and missing mass spectrum measured in $pp{\to}K^+X$ reaction.}
\label{correl3}
\end{figure}

The resulting best fit parameters are shown in Fig.~\ref{correl3}
as a function of the ratio  $|{\cal M}_t|^2/|{\cal M}_s|^2$.
The
variation of the ratio  $|{\cal M}_t|^2/|{\cal M}_s|^2$
causes rather small 
\begin{table*}[t]
\begin{center}
\caption{Five-parameter fit results for $a_s$, $r_s$,
$a_t$, $r_t$ and $|{\cal M}_s|^2$
and different ratios $|{\cal M}_t|^2/|{\cal M}_s|^2$ of the triplet and
singlet production matrix elements.}
\label{tab1}
\begin{tabular}{cccccccc}
\hline\noalign{\smallskip}
$|{\cal M}_t|^2/|{\cal M}_s|^2$&$ |{\cal M}_s|^2$ (b/sr) &
$a_s$   (fm) & $r_s$ (fm) &
$a_t$ (fm) & $r_t$ (fm) & $\chi^2$
& $\chi^2/ndf$\\
\noalign{\smallskip}\hline\noalign{\smallskip}
0.00 & $61.4_{-6.3}^{+5.9}$ & 
$-2.6_{-0.2}^{+0.2}$ & $2.47_{-0.24}^{+0.23}$ & 
$-1.5_{-0.3}^{+0.2}$ & $3.6_{-0.9}^{+0.9}$ &
29.0  & 0.805\\
0.10 & $46.7_{-4.5}^{+4.2}$ & 
$-2.7_{-0.4}^{+0.3}$ & $2.25_{-0.24}^{+0.23}$ & 
$-1.5_{-0.3}^{+0.3}$ & $3.8_{-1.0}^{+1.0}$ &
29.0 & 0.807\\
0.25 & $34.4_{-3.2}^{+3.0}$ & 
$-2.9_{-0.5}^{+0.3}$ & $1.97_{-0.23}^{+0.22}$ & 
$-1.4_{-0.4}^{+0.3}$ & $4.0_{-1.2}^{+1.1}$ &
29.1 & 0.809\\
0.50 & $24.0_{-2.3}^{+2.1}$ & 
$-3.1_{-0.6}^{+0.4}$ & $1.63_{-0.19}^{+0.18}$ & 
$-1.3_{-0.4}^{+0.4}$ & $4.5_{-1.3}^{+1.3}$ &
29.2 & 0.812\\
1.00 & $15.0_{-1.6}^{+1.4}$ &
$-3.2_{-0.6}^{+0.4} $ & $1.25_{-0.15}^{+0.13}$ &
$-1.3_{-0.5}^{+0.4}$ & $5.4_{-1.}^{+1.6}$ &
29.4 & 0.816\\
2.00 & $8.7_{-1.0}^{+0.8}$ & 
$-3.3_{-0.6}^{+0.4}$ & $0.90_{-0.10}^{+0.09}$ &
$-1.2_{-0.6}^{+0.5}$ & $6.5_{-2.1}^{+2.0}$ &
29.6 & 0.821\\
4.00 & $4.7_{-0.5}^{+0.4}$ & 
$-3.3_{-0.6}^{+0.4}$ & $0.63_{-0.07}^{+0.06}$ &
$-1.2_{-0.8}^{+0.5}$ & $7.8_{-2.7}^{+2.7}$ &
29.8 & 0.827\\
8.00 & $2.4_{-0.2}^{+0.2}$ & 
$-3.4_{-0.6}^{+0.4}$ & $0.44_{-0.05}^{+0.04}$ &
$-1.2_{-1.0}^{+0.6}$ & $9.2_{-3.5}^{+3.5}$ &
29.9 & 0.832\\
\noalign{\smallskip}\hline
\end{tabular}
\end{center}
\end{table*}
variations 
of the parameters 
$a_s$, $a_t$  
and rather large variations of the parameters
$r_s$, $r_t$,
respectively.
Therefore one can deduce important constraints
on the singlet and triplet scattering lengths
and effective ranges of the low energy $\Lambda{p}$ interaction:
\begin{eqnarray}
-4.1\; {\rm fm}&<&a_s < -2.3\; {\rm fm},\; \; \; 
r_s < 2.7\; {\rm fm}, \nonumber \\
-1.8\; {\rm fm} &<&a_t < -0.6\; {\rm fm},\; \; \; 
r_t > 2.7\; {\rm fm}. 
\label{constraint}
\end{eqnarray}

\section{Tests of potential model results.}
Available  meson-exchange potential model predictions
of the S-wave $\Lambda{p}$
singlet and triplet scattering length
and effective range parameters are listed in Table~\ref{tab2}. Here the 
results from Nijmegen model are denoted as Nijm~D~\cite{fhint:nag77}, 
Nijm~F~\cite{fhint:nag79} and Nijm~a-e~\cite{fhint:sto99}.  The  NSC 
parameters are taken from  Ref.~\cite{fhint:mae89}. The parameters 
given by J\"ulich model are denoted as J\"ul~A,B~\cite{fhint:hol89} and
J\"ul~$\tilde{\rm A}$,$\tilde{\rm B}$~\cite{fhint:reu94}. The most 
recent  J\"ulich parameters are indicated as J\"ul~03~\cite{Haidenbauer}.

\begin{figure}[t]\vspace*{-3mm}
\hspace*{-3mm}\psfig{file=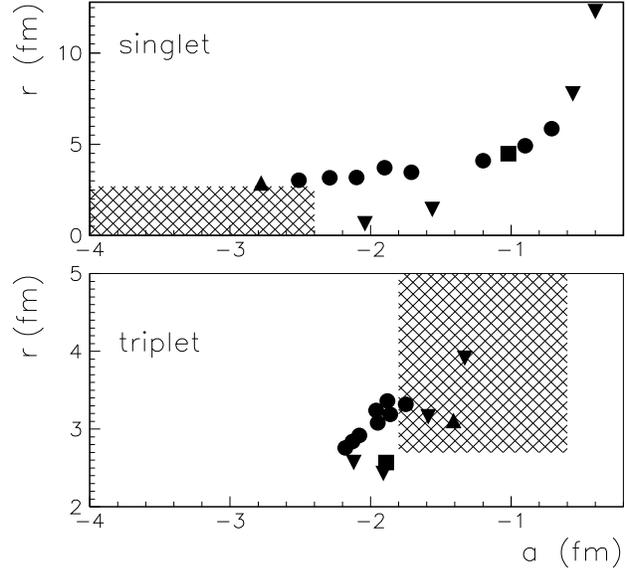,width=9.4cm,height=9.2cm}\vspace*{-3mm}
\caption{
Experimentally allowed regions (hatched rectangles) for
$\Lambda{p}$ singlet and triplet scattering lengths
and effective ranges, $a_s$, $r_s$ and $a_t$, $r_t$
deduced from five-parameter fits of the $pp\to K^+ X$
missing mass spectrum and the total $\Lambda{p}$ cross
section data. 
The symbols indicate
theoretical results for singlet and triplet parameters from different models:
circles - Nijmegen\cite{fhint:nag77,fhint:nag79}, inverse triangles -
J\"ulich~\cite{fhint:reu94,fhint:hol89} and triangle -
NSC~\cite{fhint:mae89}. The square shows most recent J\"ulich
result~\cite{Haidenbauer}.}
\label{correl2}
\end{figure}

In Fig.~\ref{correl2} the model predictions of 
$a_s$, $r_s$, $a_t$ and $r_t$ are compared with
the five-parameter fit results of Sect.~3.
The experimentally allowed regions deduced from 
five-parameter fits are indicated by hatched rectangles.
All model predictions of ($a_s$, $r_s$) and most predictions
of ($a_t$, $r_t$)
lie outside the experimentally allowed regions. 
In most cases  the singlet scattering lengths $a_s$ are too small
and the triplet scattering lengths $a_t$ are too large.


\begin{table*}[t]
\begin{center}
\caption{Test of potential model results performed with two-parameter fit.}
\label{tab2}
\begin{tabular}{lccccccccc}
\hline\noalign{\smallskip}
Model   & 
$ a_s$ \, (fm)  & $r_s$ \, (fm) &
$a_t$ \, (fm)  & $r_t$ \, (fm) & $|{\cal M}_s|^2$ \, (b/sr) &
 $|{\cal M}_t|^2$ \, (b/sr) & $\chi^2$ & $\chi^2/ndf$\\
\noalign{\smallskip}\hline\noalign{\smallskip}
Nijm a & -0.71 & 5.86 & -2.18 & 2.76&$0.\pm 0.1$&$22.8\pm 0.2$ & 47.6 & 1.22 \\
Nijm b & -0.90 & 4.92 & -2.13 & 2.84&$0.\pm 0.1$ &$23.4\pm 0.2$ & 53.9 & 1.38 \\
Nijm c & -1.20 & 4.11 & -2.08 & 2.92 &$0.\pm 0.1$&$23.9\pm 0.2$ & 62.2 & 1.60\\ 
Nijm d & -1.71 & 3.46 & -1.95 & 3.08 &$0.\pm 0.1$ &$25.0\pm 0.2$ & 83.3 & 2.14\\
Nijm e & -2.10 & 3.19 & -1.86 & 3.19 &$77.5\pm 0.7$ &$0.\pm 0.1$& 76.6 & 1.97\\
Nijm f & -2.51 & 3.03 & -1.75 & 3.32 &$74.7\pm 0.7$&$0.\pm 0.1$ & 44.8 & 1.15\\ 
J\"ul $\tilde{\rm A}$ & -2.04 & 0.64 & -1.33 & 3.91 
& $7.3\pm 0.5$& $7.8\pm 1.6$ & 55.9 & 1.43\\
J\"ul $\tilde{\rm B}$ & -0.40 & 12.28 & -2.12 & 2.57 
& $ 0. \pm 0.1$& $ 21.3\pm 0.2$ & 43.6 & 1.12\\
J\"ul A & -1.56 & 1.43 & -1.59 &3.16 &$33.8\pm 0.3$ &$0.\pm 0.1$ & 80.2&2.06\\ 
J\"ul B & -0.56 & 7.77 & -1.91 & 2.43 
& $0.\pm 0.1$ & $20.1\pm 0.2$ & 55.5 &1.42\\ 
NSC & -2.78 & 2.88 & -1.41 & 3.11 & $71.6\pm 0.7$  & $0.\pm 0.1$ & 32.8 &0.84\\ 
Nijm D & -1.90 & 3.72 & -1.96 & 3.24 
&  $ 0.\pm 0.2$  &  $ 26.1\pm 0.2$ & 91.6&2.35\\
Nijm F & -2.29 & 3.17 & -1.88 & 3.36 & $77.4\pm 0.7$ & $0.\pm 0.1$ & 62.6&1.61\\
J\"ul 03 & -1.02 & 4.49 & -1.89 & 2.57 
& $0.\pm 0.1$ & $21.3\pm 0.2$ & 62.4&1.60 \\ 
\noalign{\smallskip}\hline
\end{tabular}
\end{center}
\end{table*}

A more direct test of the model
predictions can be performed using
Eqs.(\ref{miss}) and (\ref{cross}). 
Keeping the predicted parameters fixed and fitting only the
two free parameters $|M_s|^2$ and $|M_t|^2$
the calculated cross sections are compared
with the ${\rm p}{\rm p}{\to}{\rm K}^+{\rm X}$
missing mass spectrum as well as the total $\Lambda{p}$ cross section data
and the $\chi^2$  are deduced  (see Table~\ref{tab2}).
Excepting the model J\"ul~${\tilde{\rm A}}$ this procedure yields
either $|M_s|^2{=}0$ or $|M_t|^2{=}0$.
This is due to the fact that the models predict
either a dominating singlet or triplet FSI enhancement.
The resulting values of $\chi^2$ should be compared
with the best-fit value $\chi^2{=}29.0$ obtained in direct five-parameter
fits (see Table~\ref{tab1}).
Excepting the result of the NSC model all $\chi^2$ values 
are  essentially larger than 29.0.
They indicate that those model predictions fail to
reproduce simultaneously  
the missing mass spectrum and the
total cross section data.

\section{Summary and conclusions.}

We analyzed the high resolution $pp{\to}K^+X$ 
data of Siebert et al. \cite{fhint:sie94}
with respect to the strong FSI 
near the $\Lambda{p}$ production
threshold.
The observed missing mass spectrum was described by
factorizing the reaction amplitude in terms of a production amplitude and 
FSI amplitude which was
parametrized in terms of the inverse Jost function.
It was found that a three-parameter fit with 
spin-averaged scattering length and effective range 
parameters $\bar{a}$ and $\bar{r}$ can reproduce the 
missing mass spectrum but fails to describe simultaneously
the total $\Lambda{p}$ cross section data.
Vice versa deducing $\bar{a}$ and $\bar{r}$
from a fit to the total cross section
data fails to describe the missing mass spectrum. 
Also a combined   three-parameter fit of
the mis\-sing mass spectrum and
the total $\Lambda{p}$ cross section data
fails to reproduce simultaneously both data sets 
with spin-averaged parameters $\bar{a}$ and $\bar{r}$.

Therefore the singlet and triplet scattering lengths
and effective ranges $a_s$, $r_s$, $a_t$, $r_t$
are fitted as separate free parameters.
Taking $|{\cal M}_s|^2$ as free parameter and 
the ratio $|{\cal M}_t|^2/|{\cal M}_s|^2$ 
as fixed parameter 
both the missing mass spectrum and the total
cross scetion data are perfectly reproduced
in five-parameter fits.
The ratio $|{\cal M}_t|^2/|{\cal M}_s|^2$
was varied over a wide range of values
and important  constraints on
the parameters $a_s$, $r_s$, $a_t$, $r_t$
were deduced.
These are indicated in Fig.~\ref{correl2} 
as experimentally allowed regions.

A characteristic feature of the best-fit 
solutions 
is the fact that
the $\Lambda{p}$ FSI in the reaction
$pp{\to}K^+\Lambda{p}$
is dominated by the
singlet contribution. 
Another important result 
follows from a comparison of the singlet and triplet
scattering lengths.
The fact that $-a_s > -a_t$ means that the $\Lambda{p}$ interaction
is more attractive in the singlet state than in the triplet
state. This result is in accordance with the
expectation deduced from an analysis of the
binding energies of light hypernuclei \cite{fhint:dal65}.


Though the $\chi^2$ values are only weakly dependent on
$|{\cal M}_t|^2/|{\cal M}_s|^2$ 
ratios with $|{\cal M}_t|^2/|{\cal M}_s|^2{>}8$
can be excluded on the basis of the $\chi^2$ criterion.
In this context it is interesting to note that the 
fit yields for $|{\cal M}_t|^2/|{\cal M}_s|^2{>}1$
abnormally small singlet and large triplet effective
ranges, $r_s{<}1$~fm and $r_t{>}6$~fm.
Taking the definition of the effective range  
as interaction range 
such values are questionable.
However, this problem can only  be decided by experiments
which allow to determine separately the absolute
values of the singlet and triplet
production matrix elements.

Most previous experiments provide only informations
on the spin-averaged parameters $\bar{a}$ and $\bar{r}$
and the deduced values are given without errors estimates.
But since the concept of spin-averaged parameters fails to describe
simultaneously the $pp{\to}K^+ X$ FSI enhancement  and the 
$\Lambda {p}$ total cross sections we do not compare our spin-averaged
parameters with previous determinations of $\bar{a}$ and $\bar{r}$.
In this context we mention the general
problem of defining spin-averaged parameters if singlet and triplet
parameters are different.


Inspecting the world data set 
there is one experiment which provides direct information on the
triplet parameters $a_t$ and $r_t$.
The data on the $\Lambda{p}$ production in the reaction
$K^-d{\to}\pi^-\Lambda{p}$ \cite{fhint:tai69}
show a marked FSI enhancement near the $\Lambda{p}$ threshold.
Tai Ho Tan deduced $a_t$ and $r_t$ assuming that
the $K^- d$ capture at rest occurs mainly from 
S-wave orbital admixtures.
Spin and parity considerations imply that
the final $\Lambda{p}$ state has spin one as the deuteron.
Assuming that the FSI amplitude near threshold  is dominated by the 
S-wave contribution, i.e. the $^3{\rm S}_1$-state,
the triplet scattering length and effective range
parameter $a_t$ and $r_t$ can directly be extracted
from those data. The deduced values $a_t{=}-2.0\pm 0.5$~fm
and $r_t{=}3.0\pm 1.0$~fm \cite{fhint:tai69}
agree within one standard deviation with
the experimentally allowed region
of ($a_t, r_t$). They favor  our five-parameter
solutions for small ratios $|{\cal M}_t|^2/|{\cal M}_s|^2$.
It should be mentioned that a reanalysis of the
$K^-d{\to}\pi^-\Lambda{p}$ data is highly wanted
in view of the importance and the high quality of those data.


The meson theoretical model predictions 
of the scattering lengths and effective ranges were compared with 
the experimentally allowed regions of ($a_s$, $r_s$)
and ($a_t$, $r_t$) deduced from five-parameter fits of the data.
In most cases the predicted singlet scattering lengths are too small and
the triplet scattering lengths are too large.
Only one model prediction is in agreement with the
constraints deduced from the data.
This finding was confirmed by a more direct test of the
model predictions. 
The test was performed 
by keeping  the predicted parameters in the fit routine fixed
and fitting only $|M_s|^2$ and $|M_t|^2$. 
With one exception the model predictions fail to
reproduce simultaneously the
missing mass spectrum near the $\Lambda{p}$ threshold
and the total $\Lambda{p}$  cross section data.

It should be emphasized that the shape of the FSI enhancement
near the $\Lambda{p}$ threshold is the essential experimental
information in order to evaluate the FSI parameters
from the reaction $pp{\to}K^+ \Lambda{p}$.
Since the  most important part of the FSI enhancement is
located in the sharply rising part of the missing mass spectrum
a high missing mass resolution and sufficient statistical
accuracy are needed. The effective missing mass resolution
of the SATURNE experiment \cite{fhint:sie94} corresponded
to a $1\sigma$ width of 2~MeV. 
Improving this value by an order of magnitude would allow to
study the $\Lambda{p}$ FSI  near $q{=}0$ with higher accuracy.
But not only the improvement
of precision but also a systematic check
of the method are important. An essential assumption of
the factorization ansatz (Eq.~\ref{watson}) is that
the production matrix element is 
constant, i.e. does not depend  on the internal energy
of the $\Lambda{p}$ final state. A direct check is to
measure the reaction $pp{\to}K^+ \Lambda{p}$
at different energies and angles. 


Ultimately, measurements with
polarized beams and polarized targets are needed in order to
disentangle the spin singlet and triplet contributions
in the $pp{\to}K^+\Lambda{p}$ reaction.
Such measurements would allow to determine separately
the singlet and triplet scattering lengths and effective ranges.
The feasibility depends not only on  the
polarization of beam and target but also on the
effective missing mass resolution and luminosity.

We appreciate discussions with J. Haidenbauer, C. Hanhart and H. Rohdje\ss{}.
During the preparation of the paper we were informed
about a  
theoretical study \cite{Gasparian3} of
the $pp{\to}K^+\Lambda{p}$ reaction
using dispersion relations.





\end{document}